\documentclass[preprint,12pt]{elsarticle}
\usepackage{graphicx}
\usepackage{bm}
\usepackage{epsfig}
\usepackage{amssymb}
\usepackage{amsfonts}
\usepackage{amsmath}
\usepackage{mathtools}
\usepackage{braket}
\usepackage{color}
\usepackage{epstopdf}
\usepackage{hyperref}
\usepackage{float}
\restylefloat{table}
\usepackage{bibentry}
\usepackage{multirow}
\usepackage[caption=false]{subfig}

\usepackage[normalem]{ulem} 

\usepackage{algorithm, algpseudocode}

\graphicspath{{figures/}}

\newcommand{\ba}{\begin{eqnarray}}
\newcommand{\ea}{\end{eqnarray}}
\newcommand{\bd}{\begin{displaymath}}

\newcommand{\nn}{\nonumber \\}

\newcommand{\argmax}{\mathop{\rm arg~max}\limits}

\begin{document}

\begin{frontmatter}
\title{Tensor-Ring Decomposition with Index-Splitting}

\author[a]{Hyun-Yong Lee\corref{author1}}
\author[a]{Naoki Kawashima\corref{author2}}

\cortext[author1] {Corresponding author.\\\textit{E-mail address:} hyunyong@issp.u-tokyo.ac.jp}
\cortext[author2] {Corresponding author.\\\textit{E-mail address:} kawashima@issp.u-tokyo.ac.jp}
\address[a]{Institute for Solid State Physics, University of Tokyo, Kashiwa, Chiba 277-8581, Japan}

\begin{abstract}
 Tensor-ring decomposition of tensors plays a key role in
  various applications of tensor network representation
  in physics as well as in other fields.
  In most heuristic algorithms for the tensor-ring decomposition,
  one encounters the problem of local-minima trapping.
  Particularly, the minima related to the topological structure in
  the correlation are hard to escape.
  Therefore, identification of the correlation structure,
  somewhat analogous to finding matching ends of entangled strings,
  is the task of central importance. 
  We show how this problem naturally arises in physical applications,
  and present a strategy for winning this ``string-pull'' game.
\end{abstract}


\end{frontmatter}


\section{Introduction}

Whereas the method of the tensor network can be dated back to 1960s,\cite{baxter68} when Baxter used it for numerical computation of the partition function of the dimer model on the square lattice, it is only quite recently that the tensor network has got the full attention it deserves in the physics community. Now the tensor networks are heavily used not only as a tool for achieving extremely accurate computation of model systems but also as an essential framework for conceptual developments. The use of the tensor network as the machine learning scheme has also been discussed recently.\cite{miles16}
One prominent example for the use of the tensor network is 
the real space renormalization group.
While the conventional Migdal-Kadanoff (MK) renormalization group makes
the essential idea clear, it is well-known that
the simple implementation of the idea does not yield the
correct critical behavior.
Moreover, it is hard to generalize the MK scheme so that
the accuracy is controllable and the method can represent
the exact fixed point in some limit.
An alternative real-space renormalization group method,
{\it tensor renormalization group} (TRG)\cite{levin07}
and
{\it high-order tensor renormalization group} (HOTRG)\cite{tao12}
were proposed based on the tensor network representation.

The tensor-network-based methods are not so unlike the MK scheme in essence.
In both the RG transformations, the model and the lattice is
deformed by tracing out some local degrees of freedom
to recover the same lattice structure as before but with an
enlarged lattice constant.
The difference is that we lose accuracy
in the MK scheme because of approximations we make,
the deformation in the new scheme is exact as long as
a control parameter, {\it the bond dimension}, is large enough.
Here, the bond dimension is the cut-off introduced in the truncation of
the singular values in the singular value decomposition (SVD).
Exact decomposition is indeed the case with a few initial steps
in TRG where the renormalized tensor can be split into
two sub-tensors by SVD with a relatively stringent cut-off.
Of course, as we proceed in the RG procedure,
the renormalized tensor represents a large-scale object
that carries more mutual-information, 
and can no longer be split without error.
However, in a typical application, the tensor converges to
a so-called {\it corner-double-line} (CDL) tensor,
in which the singular values decays exponentially
as a function of the index when ordered according to its magnitude,
so that the elimination of small singular values can
be an extremely good approximation.
For this reason, we can often obtain the estimates of
critical exponents with many significant digits
even when we use the bond dimension of the order of 10.

Even so, with a typical computational environment,
we cannot afford large bond dimensions, e.g., 100
in the TRG of a two-dimensional lattice model.
A similar or even severer limitation applies to
other schemes and higher dimensions.
Therefore, we have a good reason to be stingy
about the use of the bond dimension.
One problem about this is a short-range correlation.
It was pointed out\cite{gu09}
that the renormalized tensor in TRG does not converge to a {\it fixed-point tensor},
the tensor that only carries information of the fixed point.
This is because the procedure in the TRG scheme does not eliminate the correlation that lives in the scale smaller than the
renormalization scale.
The mutual information arising from this short-range correlation eats up
the capacity of the tensor, and reduces the representing
power for the fixed point properties.
This problem was solved\cite{gu09,evenbly15,shuo17,hauru18}
by introducing an additional procedure in the RG step
that explicitly eliminates local entanglement-loops
that should not affect the properties in the larger scales.
Another approach was proposed in Ref.\,\cite{harada18}
that makes the flow of the local correlation branch out
and form themselves into loops,
which are traced out and become harmless.

In either way, the essence here is the identification of the
local correlation/entanglement.
The previous algorithm that solved the problem of local entanglement loops
employed either filtering operators that eliminate the short-range correlation
or branching operators that separate the short- and long-range correlations.
They are computed iteratively with conditions that
implicitly force the resulting operator to have the desired function.
In the present article, we propose an alternative more direct method
assuming that the target tensor approximately has the CDL structure.
With this assumption on the tensor structure,
the task is reduced to something analogous to the ``string-pull'' lottery,
i.e., detecting which end is connected to which.

It is known that this structure is characteristic to
the tensors that one encounters in the renormalization group
methods based on the tensor network representation.
In what follows, we show that the CDL structure emerges
in the tensor renormalization group procedure and that
it causes difficulty in the tensor-ring decomposition (TRD) when 
previously known heuristic algorithms are applied.
For example, alternating least squares (ALS), which was originally proposed for the tensor train decomposition\,\cite{holtz12,rohwedder13}, is used for
obtaining a tensor-ring decomposition.\cite{zhao16}
While the problem is not so prominent when a generic tensor
is considered as the target tensor,
its application to the CDL tensors reveals the drawback
of previous iterative approaches.
In this article, we show that it is indeed the case and
that the new algorithm produces the optimal branching tensors
within a much small number of iterations.

\begin{figure}[h]
	\centering
  \includegraphics[width=0.5\textwidth]{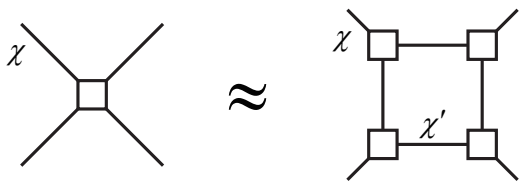}
  \caption{Tensor-ring decomposition of a tensor of the fourth-order. }
  \label{fig:TRD}
\end{figure}

In the present article, we focus on the TRD of tensors.
A tensor ring is a partially contracted set of tensors arrayed on a ring.
For instance, if a tensor, $T_{ijkl}$, of the fourth-order is given,
we may consider four tensors of the third-order that satisfy
\begin{eqnarray}
  T_{ijkl} \approx
  \sum_{\alpha\beta\gamma\delta} Z^1_{i\alpha\beta} Z^2_{j\beta\gamma} Z^3_{k\gamma\delta} Z^4_{l\delta\alpha}.	
\end{eqnarray}
The right-hand side is an example of the tensor ring.
The TRD is schematically depicted in Fig.\ref{fig:TRD}.
It can be shown that we can exactly express any tensor
in terms of a tensor ring if we allow each contracted index
to have large enough dimension.
In the case of the example shown above,
where the original tensor have four indices ($i,j,k$ and $l$)
of dimension $d$,
we can have exact mapping if
the representation can be exact if each one of the contracted
indices, $\alpha,\beta,\gamma$, and $\delta$, has the dimension
$\chi \ge d^2$.
However, it is often desirable to obtain a tensor ring
that is an accurate representation, though not exact,
while the dimension of the inner bonds are small enough to
meet some practical restrictions.
Although it is not so widely appreciated so far,
the TRD is a key procedure
for various numerical calculations.
In the RG calculation, for example,
the TRD can be used as an alternative to the
loop filtering in the loop tensor network renormalization proposed in Ref.\,\cite{shuo17}.

\section{CDL and HOSVD}

\begin{figure}[h]
  \begin{center}
    \includegraphics[width=0.6\textwidth]{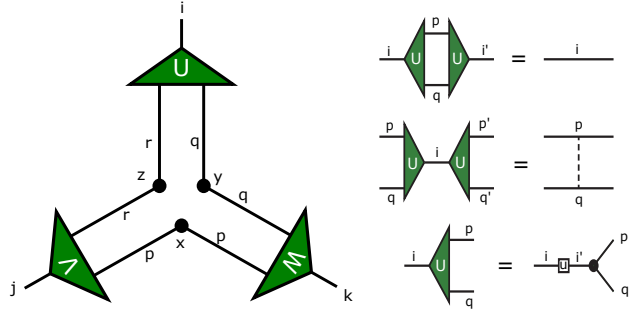}
    \caption{The target tensor is a pseudo CDL.}
    \label{fig:CDL}
  \end{center}
\end{figure}
In what follows, we mainly consider the case where the target tensor is
of the third-order because it is the simplest non-trivial case
and the problem with higher orders can be solved by repeating use of
the method for the third-order tensors.
Also, we focus on the {\it pseudo-CDL} tensors as defined below.
A pseudo-CDL tensor is a tensor that can be expressed in terms of 
{\it pseudo-unitary} matrices ($U, V$ and $W$)
and positive weights $x_p,y_q,z_r\ge 0$ as
\begin{equation}
  T_{ijk} = \sum_{pqr} U_{iqr} V_{jrp} W_{kpq} x_p y_q z_r.
  \label{eq:pCDL}
\end{equation}
Here, a pseudo-unitary matrix is an isometry that satisfies
\begin{equation}
  \sum_{qr} U_{iqr} U^{\ast}_{i'qr} = \delta_{ii'} \ \mbox{(isometry)},
  \label{eq:Isometry}
\end{equation}
and
\begin{equation}
  \sum_{i} U_{iqr} U^{\ast}_{iq'r'} = \delta_{qq'} \delta_{rr'} \Delta_{qr},
  \label{eq:PseudoUnitary}
\end{equation}
with some function $\Delta_{qr} = 0$ or $1$.
Diagrammatically, $T_{ijk}$ is represented as Fig.\ref{fig:CDL}.
We call a tensor $T_{ijk}$ a pseudo-CDL tensor if
it is expressed in the form of Eq.\,(\ref{eq:pCDL})
with Eq.\,(\ref{eq:Isometry}) and Eq.\,(\ref{eq:PseudoUnitary}).
The pseudo-CDL tensor $T_{ijk}$ is a full CDL tensor
if $U,V$, and $W$ are unitary (i.e., $\Delta_{qr} = 1$ for $U$).
In what follows, we refer a pseudo-CDL simply as a CDL.
We impose an additional condition to remove redundancy in the representation
that the indices are defined so that weight is sorted in the descending order
$x_0 > x_1 > \cdots > 0$.

Note that we can express the isometries $U$
by using a mapping from subindex pairs to indices.
To be more specific, let us consider the sets of the values
that $q$, $r$, and $i$ take, i.e.,
$\Omega_q=\{1,2,\cdots,d_q\}$, $\Omega_r=\{1,2,\cdots,d_r\}$,
and $\Omega_i=\{1,2,\cdots,D_i\}$ where $d_q,d_r$ and $D_i$ are
dimensions of indices $q,r$ and $i$, respectively.
Then, we consider an ``indexing function'', $I(q,r)$, from
$\Omega_q \times \Omega_r$ to $\Omega_i\cup \{0\}$.
We assume that $I$ is bijective when it is restricted to $I^{-1}(\Omega_i)$.
Using this function we can define a third-order tensor
\begin{equation}
  \Delta_{iqr}^U \equiv \delta_{i,I(q,r)}. 
  \label{eq:IndexSplitter}
\end{equation}
We can prove that if both the conditions
Eqs.\,(\ref{eq:Isometry}) and (\ref{eq:PseudoUnitary} )hold,
then such a (partially) bijective function $I(q,r)$,
and therefore the tensor $\Delta_{iqr}^U$ as well,
exist and the isometry $U$ can be expressed simply as
a unitary rotation applied to the first index of $\Delta_{iqr}^U$, i.e.,
\begin{equation}
  U_{iqr} = \sum_{i'=1}^{D_i} u_{ii'} \Delta_{i'qr}^U,
  \label{eq:PseudoUnitaryStandard}
\end{equation}
where $u_{ii'}$ is a unitary matrix.

If we know $U,V,W,x,y$ and $z$, the TRD is given as
\begin{equation}
  T_{ijk} = \sum_{pqr} Z^1_{iqr} Z^2_{jrp} Z^3_{kpq}
\end{equation}
where the latent tensors
\begin{eqnarray}
  Z^1_{iqr} = U_{iqr} \sqrt{y_q z_r}, \nn
  Z^2_{jrp} = V_{jrp} \sqrt{z_r x_p}, \nn
  Z^3_{kpq} = W_{kpq} \sqrt{x_p y_q}.
	\label{eq:exact_decomp}	
\end{eqnarray}
Therefore, our problem is to explicitly compute
$U,V,W,x,y$ and $z$ for a given $T_{ijk}$.

To this end, we start from the high-order singular value decomposition (HOSVD).
It is known\,\cite{tucker66} that an arbitrary tensor (a third-order tensor in
the present case) can be expressed as
\begin{equation}
  T_{ijk} = \sum_{i'j'k'} u_{ii'} v_{jj'} w_{kk'} t_{i'j'k'}, \label{eq:HOSVD}
\end{equation}
where the matrices $u,v$ and $w$ are unitaries and $t_{ijk}$ is a tensor whose 
matrix slices are mutually orthogonal, e.g.,
\begin{equation}
  \sum_{jk} t_{ijk} t_{i'jk}^{\ast} = a_i \delta_{ii'}, \, \cdots.
  \label{eq:MutuallyOrthogonal}
\end{equation}
In fact, we can obtain the decomposition of Eq.\,(\ref{eq:HOSVD}) by
applying the SVD repeatedly for each index, i.e.,
\begin{eqnarray}
  T_{ijk} & = \sum_{i'} u_{ii'} s^1_{i'} \bar t^1_{i'jk}, \nonumber\\
  s^1_{i'} \bar t^1_{i'jk} & = \sum_{j'} v_{jj'} s^2_{j'} \bar t^2_{i'j'k}, \nonumber\\
  s^2_{j'} \bar t^2_{i'j'k} & = \sum_{k'} w_{kk'} s^3_{k'} \bar t^3_{i'j'k'},
  \label{eq:svd_repeat}
\end{eqnarray}
and
\begin{eqnarray}
  t_{ijk} \equiv s^3_{k} \bar w_{ijk}.
\end{eqnarray}
We call $t_{ijk}$ the ``core'' tensor.
Here, again, to remove redundancy, we require that the indices are
defined so that $s^{\lambda}_0 > s^{\lambda}_1 > \cdots > 0$ $(\lambda=1,2,3)$.
With this restriction, the decomposition (\ref{eq:HOSVD}) is
unique up to the gauge degrees of freedom\cite{lathauwer00}, i.e.,
\begin{eqnarray}
	& t_{ijk} \rightarrow e^{i(\alpha_i+\beta_j+\gamma_k)} t_{ijk},\quad u_{ii'} \rightarrow e^{-i\alpha_{i'}} u_{ii'}, \nn
	& v_{jj'} \rightarrow e^{-i\beta_{j'}} v_{jj'}, \quad
	  w_{kk'} \rightarrow e^{-i\gamma_{k'}} w_{kk'}.
  \label{eq:gauge_freedom}	
\end{eqnarray}
Note that Eq.\,(\ref{eq:pCDL}) can be cast into the form of Eq.\,(\ref{eq:HOSVD}),
by expressing the $U,V$ and $W$ as in Eq.\,(\ref{eq:PseudoUnitaryStandard}).
Namely,
\begin{equation}
  T_{ijk} = \sum_{i'j'k'} u_{ii'} v_{jj'} w_{kk'} t'_{i'j'k'}
  \label{eq:HOSVDofCDL}
\end{equation}
where
\begin{equation}
  t'_{ijk} \equiv
  \sum_{pqr} \Delta^U_{iqr} \Delta^V_{jrp} \Delta^W_{kpq} x_p y_q z_r
  \label{eq:StandardForm}
\end{equation}
Here the ``indexing'' tensors $\Delta^a$\,($a=U,V$ and $W$) are defined by
using some indexing functions $I(q,r)$, $J(r,p)$ and $K(p,q)$, respectively, as
\begin{equation}
  \Delta^U_{iqr} \equiv \delta_{i,I(q,r)}, \,\,\,\,
  \Delta^V_{jrp} \equiv \delta_{j,J(r,p)}, \,\,\,\,
  \Delta^W_{kpq} \equiv \delta_{k,K(p,q)}.
\end{equation}
We can easily verify that, because of Eq.\,(\ref{eq:PseudoUnitary}),
$t'_{ijk}$'s matrix slices are mutually orthogonal, i.e., $t'_{ijk}$ satisfies
Eq.\,(\ref{eq:MutuallyOrthogonal}).
This means that Eq.\,(\ref{eq:HOSVDofCDL}) is a HOSVD.
The uniqueness of the HOSVD, then, demands that
$t'$ is identical to the core tensor of HOSVD,
apart from the order of the indices and the gauge factors.
This is an important observation in computing $U,V,W$
and $x,y,z$ in the following section.

\begin{figure}[t]
	\centering
  \includegraphics[width=0.6\textwidth]{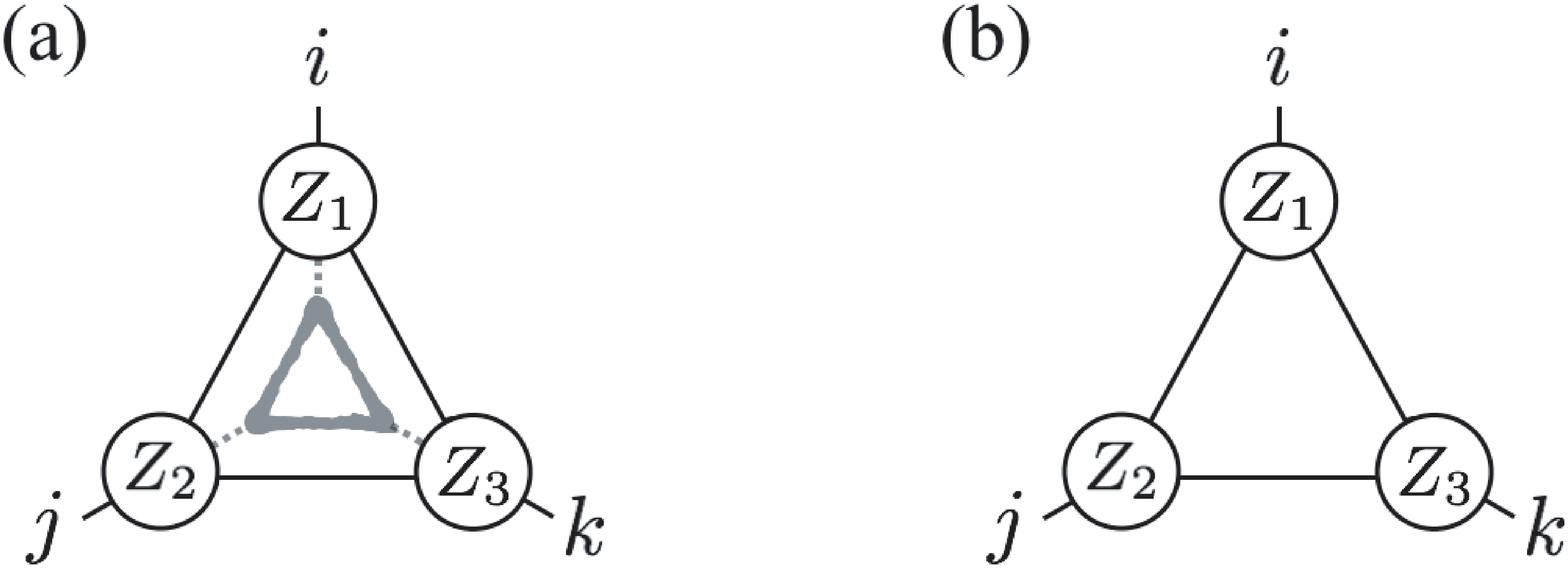}
  \caption{ Schematic figure }
  \label{fig:three_leg_cdl}
\end{figure}

One can also decompose the tensor $T_{ijk}$\,[Eq.\,(\ref{eq:pCDL})] into three latent tensors $\{Z^1,Z^2,Z^3\}$ by applying previously proposed algorithms such as the sequential SVD or ALS and its applied versions in Ref.\,\cite{zhao16}:

\begin{eqnarray}
	T_{ijk} \stackrel{\rm TRD}{\longrightarrow} \sum_{pqr} Z^1_{iqr} Z^2_{jrp} Z^3_{kpq}.
\end{eqnarray}
However, it is obvious that such decompositions do not remove
the so-called {\it short-range entanglement} loop inside
the ring network\,[gray line in Fig.\,\ref{fig:three_leg_cdl}\,(b)].
For example, in the sequential SVD method,
we start with splitting an index, which connects the first and rest
of latent tensors,
into two subindices without concerning such indexing function.
Then, it gives rise to an arbitrary entanglement by mixing the weight vectors.
Also, in ALS algorithm, we initialize the latent tensors as random tensors,
which are generally known to be {\it maximally} entangled\,\cite{hayden04}.
Optimization may discard many of such unexpected entanglements introduced at the beginning of both algorithms.
However, some local entanglement may remain as a redundancy of the
resulting tensor-ring representation.
For example, consider the latent tensor of the form
\begin{equation}
  Z^{\mu}_{iqr} = A^{\mu}_{iq_1r_1} \delta_{q_2r_2}
\end{equation}
where $q_1$ and $q_2$ are subindices so that there is one-to-one
correspondence between $q$ and $(q_1,q_2)$.
Then,
\begin{align}
  \sum_{pqr}
  Z^{1}_{iqr} Z^{2}_{irp} Z^{3}_{ipq}
  & =
  \sum_{pq_1q_2r_1r_2}
  A^{1}_{iq_1r_1} A^{2}_{ir_1p_1} A^{3}_{ip_1q_1}
  \delta_{q_2r_2} \delta_{r_2p_2} \delta_{p_2q_2}\nn
  & \propto
  \sum_{pq_1r_1}
  A^{1}_{iq_1r_1} A^{2}_{ir_1p_1} A^{3}_{ip_1q_1}.	
\end{align}
In this case, it is obvious that we can simply replace
$Z$ by a more compact tensor $A$ without losing anything,
while detecting such a redundancy is not trivial.
An example is symbolically depicted in Fig.\,\ref{fig:three_leg_cdl}(a).
Such redundancy is not only unnecessary
for describing $T_{ijk}$ but also leads to
a higher computational complexity with less accuracy
by requiring a larger bond dimension.
We call this type of redundancy an {\it entanglement loop}.
On the contrary, an ideal decomposition provided
in Eq.\,(\ref{eq:exact_decomp}) would be like
Fig.\,\ref{fig:three_leg_cdl}\,(b) where
the short-range entanglement is completely discarded.
%

\section{Algorithm for Pure CDL}

In this section, we propose a systematic way to find the ideal latent tensors\,[Eq.\,(\ref{eq:exact_decomp})] for TRD of CDL tensors. To this end, our main task is to find $U,V,W,x,y$ and $z$ for a given $T_{ijk}$ as mentioned earlier. The whole procedure consists of four steps:

\begin{enumerate}
	\item Finding the weight vectors $x_p,y_q$ and $z_r$
	\item Finding the indexing tensors $\Delta_{iqr}^U$, $\Delta_{jrp}^V$ and $\Delta_{kpq}^W$
	\item Fixing the gauge freedom in Eq.\,(\ref{eq:gauge_freedom})
	\item Constructing TRD using the obtained weight vectors, indexing tensors and gauge factors
\end{enumerate}
Obtaining the indexing tensors and fixing the gauge, one can construct the isometries $U,V$ and $W$ up to the gauge. In the following, we present the details of each step.

\subsection{Weight vectors}

First, we decompose the tensor $T_{ijk}$ into unitaries\,($u_{ii'},v_{jj'}$ and $w_{kk'}$)
and a core tensor\,($t_{ijk}$) by applying HOSVD\,[Eq.\,(\ref{eq:HOSVD})]:

\begin{eqnarray}
	\includegraphics[width=0.5\textwidth]{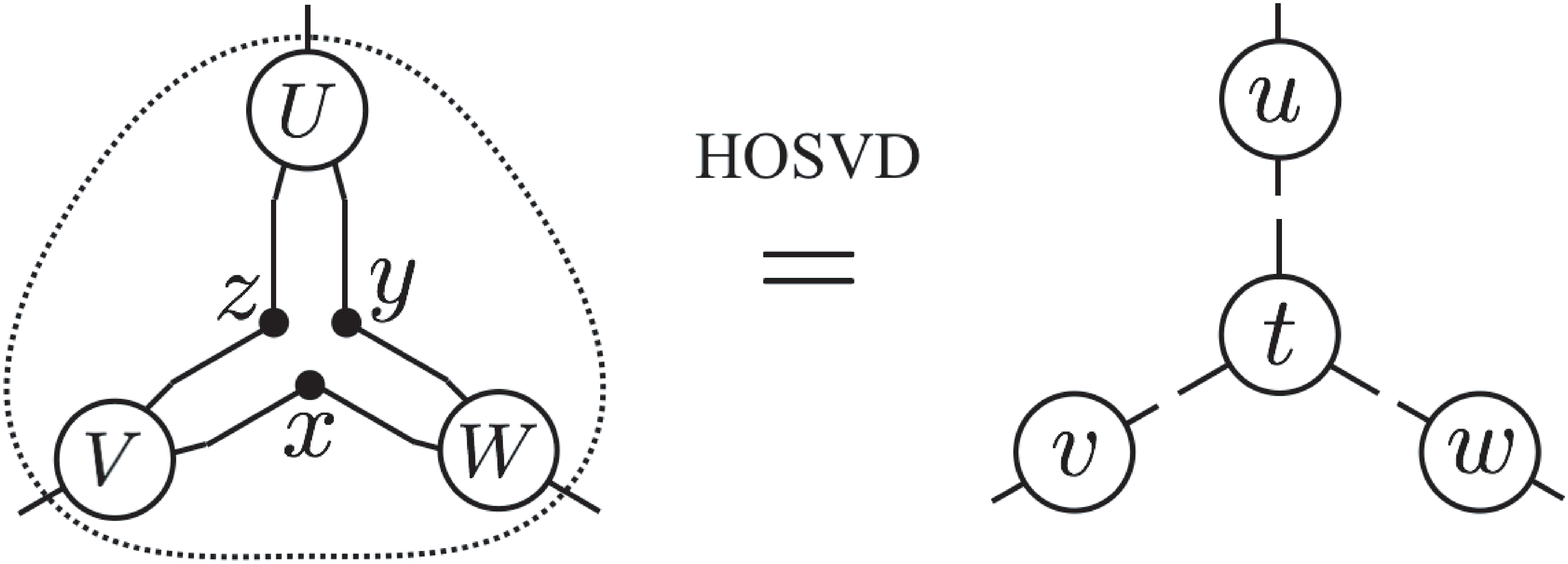}.
	\label{eq:first_hosvd}	
\end{eqnarray}
To avoid confusion, we enclose the tensor network with a dotted loop throughout this paper if its internal structure is unknown, such as the left-hand side of the above equation. As pointed out in the previous section, the tensor $U$ is similar to $u$ up to the random phase difference and order of each column vector: $U_{iqr} = \sum_{i'} u_{ii'} \Phi^U_{i'} \Delta^U_{i'qr}$ where $\Phi^U_{i'} = e^{i\alpha_{i'}}$\,(similar for $V$ and $W$).  
Due to the gauge redundancy, the core tensor $t_{ijk}$ contains, in general, complex elements, and  
its internal structure of $t$ is the following: 

\begin{eqnarray}
	t_{ijk} = &\sum_{pqr} (\Phi^U_i)^*(\Phi^V_j)^*(\Phi^W_k)^* \Delta_{iqr}^U \Delta_{jrp}^V  \Delta_{kpq}^W x_p y_q z_r
\end{eqnarray}
or

\begin{eqnarray}
	\includegraphics[width=0.4\textwidth]{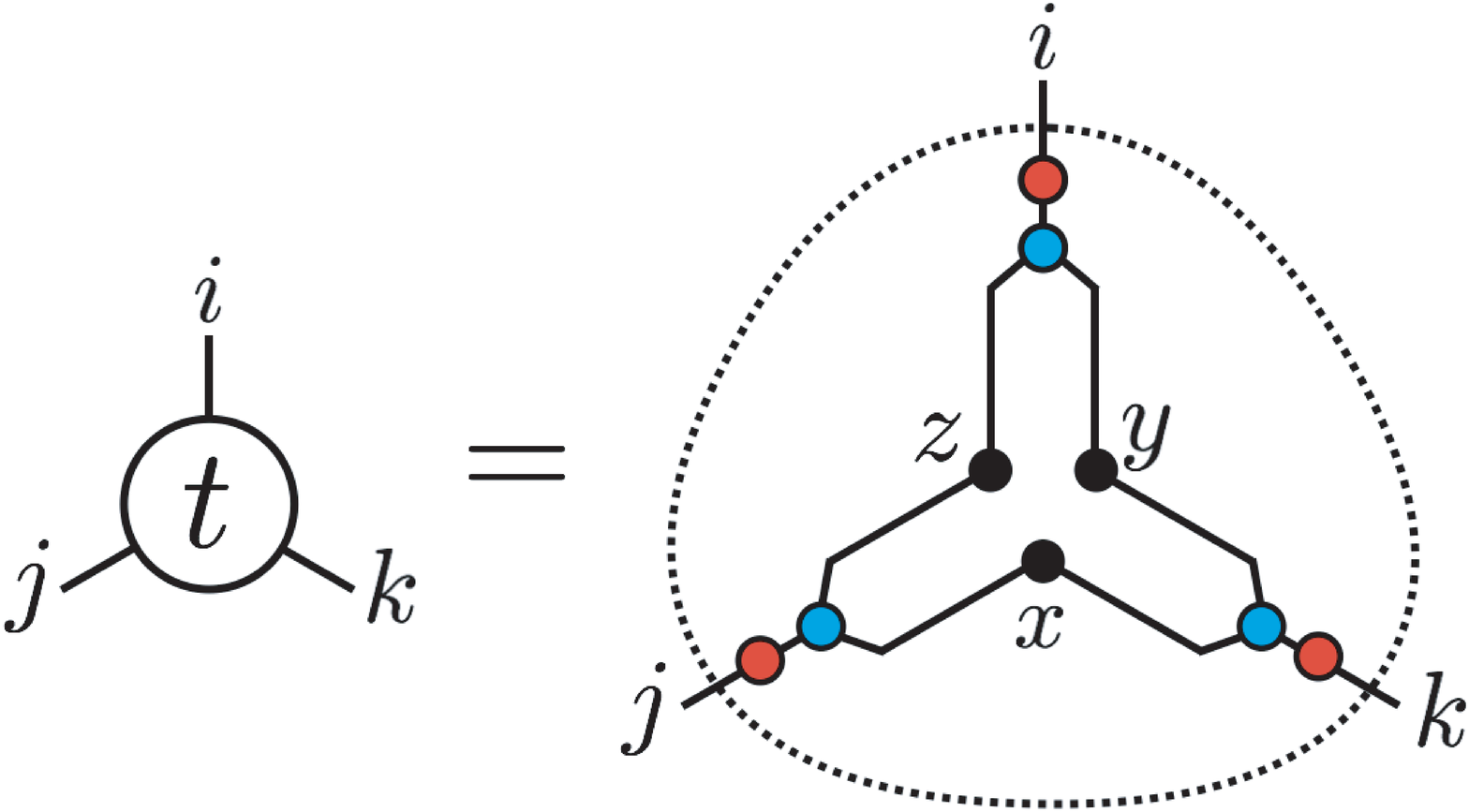},
	\label{eq:abs_C}
\end{eqnarray}
where each blue\,(red) circle denotes $\Delta^a\,(\Phi^a)$ and $a=U,V$ and $W$.
Since we are assuming those weights to be real and positive,
eliminating the phases $\Phi^x$ by taking the absolute value of $t$-tensor
does not affect them: $|t|_{ijk}$.
Then, one can extract the $x_p$, $y_q$ and $z_r$ from $|t|_{ijk}$ as follows.
Note that, because of the special internal structure of the CDL tensor,
the $|t|$-tensor has the following property.
When one of the three indices, say $i$, is fixed to be $i_0$, there is a one-to-one correspondence between non-zero elements of $|t|_{i_0jk}$ and the subindex weight $x_p$. Therefore, the list obtained by sorting the non-zero elements of $|t|_{i_0 j k}$ in descending order must be identical to $x_p$ up to an overall constant\,($c$):
\begin{equation}
    \underset{j,k \,\mbox{:}\, t_{i_0jk}\ne 0}{\mathcal Sort} \left( |t_{i_0jk}|, \, `descend' \right)
    = c\times (x_1, x_2, \cdots), \label{eq:cdl_m2_new}
\end{equation}
where the function ${\mathcal Sort}$ returns the list of the values of the first argument sorted in descending order.
Also, one can easily verity that the constant $c$ is always unity for $i_0=1$,
resulting in $\tilde{x} = x$.
In a similar way, by fixing $j$- and $k$-indices to $1$,
one can find respectively $y$ and $z$, and then our first task is done. 

\subsection{Indexing tensors}

Next, let us find the indexing tensors $\Delta^a_{iqr}$
which is supposed to sort the singular values
$s_i^a$\,[in Eq.\,(\ref{eq:svd_repeat})] in descending order.
Again we will consider the case where $a=`U'$, and the extension to the
other cases should be straight-forward.
From the internal structure of $T_{ijk}$ in Fig.\,\ref{fig:CDL},
one can easily see that the $s_i^U$ is simply a vector obtained
by Kronecker product of two weight vectors $y$ and $z$.
Then, our index function $I(q,r)$ is such that 
we can find the element $y_q z_r$ in the $I(q,r)$-th place in the sorted list
\begin{equation}
  y_{q_1} z_{r_1} >  y_{q_2} z_{r_2} >  y_{q_3} z_{r_3} > \cdots > y_{q_{D_i}} z_{r_{D_i}}. 
\end{equation}
(When the dimension of the index is smaller than the product of
the dimensions of $q$ and $r$, we must truncate the list and
assign $I(q,r)=0$ to the elements beyond the dimension of the index $i$, i.e., $D_i$.)
It is straight-forward to find such a function once we obtain the vectors $y$ and $z$.
%
%
Then, we can define the $\Delta^U$ tensor as 
\begin{eqnarray}
	\Delta^U_{iqr} = \delta_{i,I(qr)}.
\end{eqnarray}
%

%

%
%
%
%
%
%

%
\subsection{Gauge factors}
\renewcommand{\algorithmicrequire}{\textbf{Input:}}
\renewcommand{\algorithmicensure}{\textbf{Output:}}
\begin{algorithm}
  \caption{Finding $\Phi^a$ }
  \label{alg:phi}
\begin{algorithmic}[1]
	\Require A $n$th-order core tensor $t$ of size $(D_{i_1}\times\cdots\times D_{i_n})$.
	\Ensure ${\Phi}^a$ with $a=1,\dots,n$ which satisfies Eq.\,(\ref{eq:phis}).
	\State Define and initialize an $n$th-order tensor $\widetilde{t}$
        with the input tensor $t$: $\widetilde{t}_{i_1 i_2\cdots i_n} = t_{i_1 i_2 \cdots i_n}$ 
	\For { $l=1$ to $n$ } 
	\For { $i_{l}'=1$ to $D_{i_l}$} 
	\State Find the maximum value\,($c$) of the tensor $t$ with 
	\State \quad fixed $l$-th index to $i_l'$ :\ 
	$c := {\rm max}\left( t_{i_1 \cdots i_{l}' \cdots i_n}, 'abs' \right)$
	\State \quad where $'abs'$ denotes the absolute maximum.  
	\State Update $\Phi$:\ $\Phi^l_{i_{l}'} := |c|/c$
	\EndFor 
	\State Update $\widetilde{t}$ as follows:
	  \begin{eqnarray}
	    \widetilde{t}_{i_1 i_2 \cdots i_n } :=
            ({\Phi}^l_{i_l})^* \widetilde{t}_{i_1 i_2 \cdots i_n}\nonumber
	  \end{eqnarray}
	\EndFor
\end{algorithmic}
\end{algorithm}

Now, let us turn to the phase factors $\Phi^a$ which should satisfy
\begin{eqnarray}
	(\Phi^U_i)^* (\Phi^V_j)^* (\Phi^W_k)^* t_{ijk} = |t|_{ijk}.
	\label{eq:phis}
\end{eqnarray}
Due to the gauge freedom pointed out in Eq.\,(\ref{eq:gauge_freedom}),
there is an infinite number of sets $\{\Phi^a\}$ respecting the above relation.
However, any choice of the gauge does not affect the resulting tensor elements.
Here, we only present an algorithm (Algorithm \ref{alg:phi}) that finds one of the solutions
and leave the proof for its validity to Appendix. 

\subsection{Tensor ring decomposition}

Using the obtained weight vectors, indexing tensors, gauge factors and unitaries from HOSVD,
one can construct the ring tensor network exactly representing the CDL, $T_{ijk}$.
As shown in Eq.\,(\ref{eq:exact_decomp}), the latent tensors are given by

\begin{eqnarray}
	\includegraphics[width=0.55\textwidth]{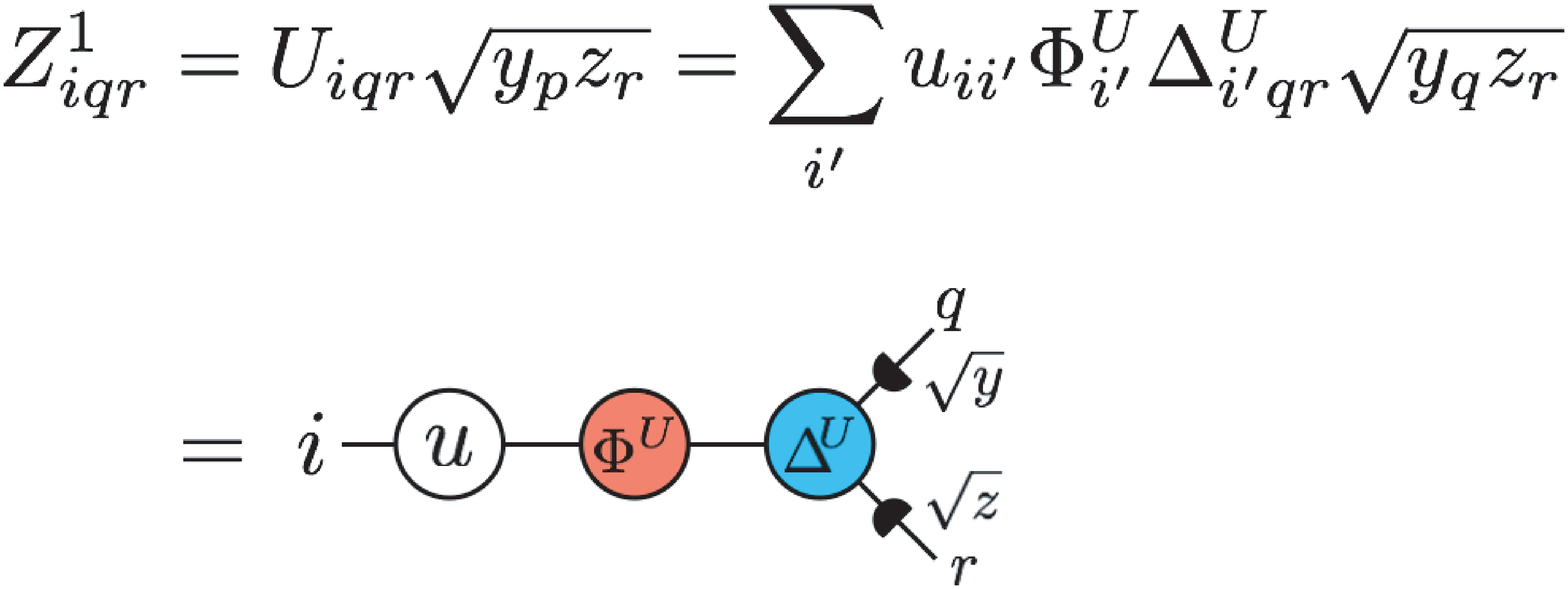},
	\label{eq:identity_insert}
\end{eqnarray}
and similarly
\begin{eqnarray}
	& Z_{jrp}^2 = \sum_{j'} v_{jj'} \Phi_{j'}^V \Delta_{j'rp}^V \sqrt{z_r x_p}, \nn
	& Z_{kpq}^3 = \sum_{k'} w_{kk'} \Phi_{k'}^W \Delta_{k'pq}^W \sqrt{x_p y_q}.
\end{eqnarray}
Finally, the TRD of the tensor $T_{I;J;K}$ is accomplished with the latent tensors $\{Z^1,Z^2,Z^3\}$:
\begin{eqnarray}
	T_{ijk} = \sum_{pqr} Z^1_{iqr} Z^2_{jrp} Z^3_{kpq}.
\end{eqnarray}
The above decomposition is exact as long as the original tensor has the exact CDL
structure\,[Fig.\,\ref{fig:three_leg_cdl}\,(a)],
and the algorithm is readily generalized to the higher-order tensors.

\subsection{Weak noise }

In practical calculations such as the tensor renormalization group method,
it is generally inevitable to encounter some noise in tensors.
Unfortunately, the algorithm introduced in the previous subsection applies to
only {\it pure} CDL tensors without any noise. In this subsection,
we discuss how the TRD using IS, referred to as IS-TRD hereafter,
can be modified to apply to such general cases.

Let us introduce random noises in the tensor as follows

\begin{eqnarray}
    T_{ijk}^{\gamma} = T_{ijk} + \Gamma_{ijk},
\end{eqnarray}
where $\Gamma_{ijk}$s are independent uniform random numbers distributed in $[-\gamma,\gamma]$.
Note that one can reasonably approximate the tensor as a pure CDL tensor as long as the noise
amplitude $\gamma$ is small enough, i.e., $\gamma \ll 1$
where the largest element of $T_{ijk}$ is set to unity:
$ T_{ijk}^{\gamma \ll 1} \simeq T_{ijk}^{\gamma=0}$.
Therefore, in such case, the IS-TRD is still valid and expected to give the best ansatz for the TRD.
However, because of the random noise,
when we extract the weight vectors $x, y$ and $z$ in Eq.\,(\ref{eq:cdl_m2_new}),
the number of non-zero elements in $|t|_{1jk}$ is not $d_p$ but $D_j D_k$.
We, therefore, should introduce a cut-off in Eq.\,(\ref{eq:cdl_m2_new}):

\begin{equation}
  x = \underset{j,k \,\mbox{:}\, t_{1jk}\ne 0}{\mathcal Sort} (\, |t|_{1jk}, \,'descend', d_p),
  \label{eq:cdl_noise_m2}
\end{equation}
where the third argument $d_p$ in ${\mathcal Sort}$ function denotes the size of
the returned list.
One can find the $y$ and $z$ vectors in a similar way,
and then find the $\Delta^a$, $\Phi^s$ and $\{Z_1,Z_2,Z_3\}$
by following the same procedure as the pure CDL case.
With a finite noise\,($\gamma > 0$), the TRD
\begin{eqnarray}
	T_{ijk}^{\gamma} \simeq \sum_{pqr} Z^1_{iqr} Z^2_{jrp} Z^3_{kpq},
\end{eqnarray}
with IS is not exact.
However, it can be an excellent approximate decomposition for weak enough noise as demonstrated below.
In addition, one can use the obtained $\{Z_1,Z_2,Z_3\}$ as a set of initial latent tensors
and apply other iterative methods, e.g., ALS algorithm, to obtain better ansatz.
In the present article, we combine the IS and ALS algorithms\,(IS-ALS for short) to obtain
the best TR decomposition for the CDL with weak noise.

\section{Benchmark}


%
\begin{figure}[ht!]
	\centering
  \includegraphics[width=0.6\textwidth]{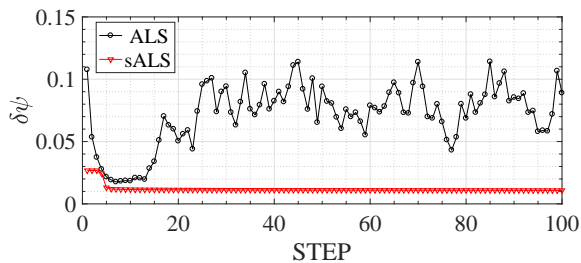}
  \caption{
    The dynamics of the approximation error $\delta\psi$ obtained by ALS
    and sALS for a rank-3 target tensor as a function of ALS step.
    The target tensor is generated by applying a random site-wise unitary
    transformation to a CDL tensor with inner-bonds of the rank 3 ($d_p=d_q=d_r=3$).
    The truncation dimension of the inner indices, or TR-rank,
    is fixed to be equal to the inner-bond rank of the target tensor,
    i.e., $\chi=\chi_p=\chi_q= \cdots =3$.
  }
  \label{fig:als_sals}
\end{figure}
\begin{figure}[ht!]
	\centering
  \includegraphics[width=0.7\textwidth]{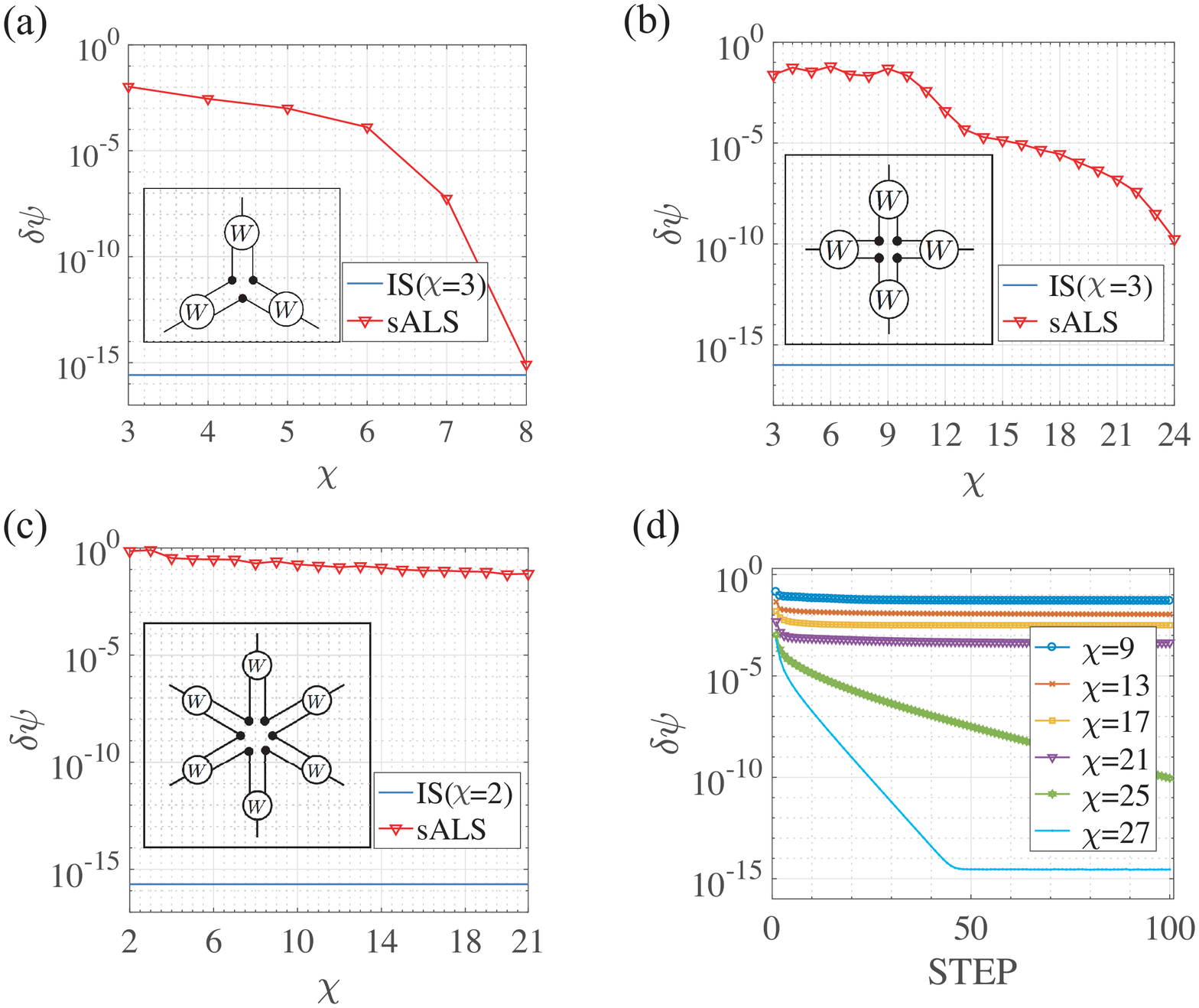}
  \caption{ The comparison of the approximation errors $\delta\psi$ obtained
    by the IS and sALS algorithms as a function of the inner-bond dimension $\chi$
    of the ansatz. The target tensors are
    (a) a third-order tensor with the dimension of the inner bonds fixed to be 3,
    i.e., $d \equiv d_p=d_q=\cdots=3$,
    (b) a fourth-roder tensor with $d \equiv d_p=d_q=\cdots=3$
    and (c) sixth-order tensor with $d=2$.
    In (d), the dynamics of $\delta\psi$ obtained by sALS for a fourth-order
    tensor with $d=3$ as a function of ALS step.}
  \label{fig:cdl_benchmark}
\end{figure}

We have found that the ALS, which optimizes the initial {\it random} latent tensors iteratively, strongly depends on the initial condition and generally fails to converge to a stable ansatz for the pure or weakly disordered CDL tensors as shown in Fig.\,\ref{fig:als_sals}\,(black circle). Here, the vertical axis $\delta \psi =\, \parallel\! T - {\rm tTr}\{Z^a\} \!\parallel_F/\parallel\!T\!\parallel_F$ stands for the approximation error by the TRD, where $\parallel \cdots \parallel_F$ and ${\rm tTr}\{ Z^a \}$ denotes the Frobenius norm and the contraction of the latent tensors $\{ Z^a \}$. The horizontal axis indicates the iteration step in the ALS algorithm. In order to make the ALS stable for the CDL tensors, we propose to set the latent tensors obtained by the sequential SVD algorithm\cite{zhao16} as the initial latent tensors of ALS\,(we call it {\it sALS} throughout this article). Even though sALS does not eliminate the unnecessary entanglement loop in the TRD, it is surprisingly stable and gives better ansatz than the ones obtained by ALS as shown in Fig.\,\ref{fig:als_sals}\,(red triangle). Therefore, we examine the performance of the proposed IS and IS-ALS algorithms by comparing them with the sALS decomposition. 

First, we present the results of TRD errors by sALS and IS for third-order,
fourth-order and sixth-order pure CDL tensors
in Fig.\,\ref{fig:cdl_benchmark}\,(a)-(c), respectively.
The horizontal axis $\chi$ denotes the so-called {\it TR-rank}\cite{zhao16},
which is the truncation dimension of the inner bonds of the trial tensor.
In other words, $\chi$ is the parameter that characterizes the capacity of
the ansatz or the trial tensor.
On the other hand, we denote the inner-bond dimension of the target tensor by $d$.
For a target tensor, the CDL weight vectors $x_p, y_q, z_r, \cdots$
and unitary matrices $U, V, W, \cdots$ are randomly chosen,
In doing so, the inner-bond dimension of the target tensor is fixed.
We use $d=3$ for the third-order and fourth-order
test tensors\,[Fig.\,\ref{fig:cdl_benchmark}\,(a) and (b)]
and $d=2$ for the sixth-order test tensor\,[Fig.\,\ref{fig:cdl_benchmark}\,(c)], respectively.
For IS, since the $\chi=d$ turned out to be sufficient for obtaining the optimal
result, $\chi$ is set to be equal to $d$ in all calculations.
As mentioned earlier, the IS decomposition is exact, i.e. $\delta \psi$ is zero
up to the machine precision $O(10^{-16})$\,
[blue solid line in Fig.\,\ref{fig:cdl_benchmark}\,(a)-(c)],
for pure CDL tensors\,($\gamma=0$).
On the other hand, the sALS algorithm fails to produce
the ideal ansatz even when $\chi$ is much larger than $d$,
which suggests that the ring decomposition obtained by the sALS algorithm
is redundant due to some inner entanglement loop, as anticipated.
Fig.\,\ref{fig:cdl_benchmark}\,(d) presents the dynamics of $\delta \psi$ obtained
by sALS for a third-order CDL tensor with $\chi=3$ as a function of the iteration step.
It indicates that the optimization gets stuck in a local minimum
and thus cannot flow into the best solution for a given $\chi$.
This also shows how efficient the IS algorithm is for the CDL tensor
without any iteration.

\begin{figure*}[!t]
  \includegraphics[width=1.0\textwidth]{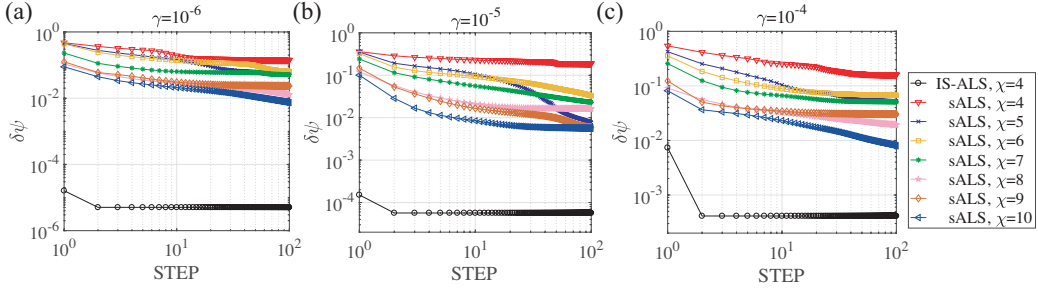}
  \caption{ Noise dependence of the performance of IS-ALS and sALS for a (randomly chosen)
    4th-degree disordered CDL tensor with the disorder amplitude
    (a) $\gamma=10^{-6}$, (b) $\gamma=10^{-5}$ and (c) $\gamma=10^{-4}$.
    The target tensors are obtained from randomly generated
    CDL with the 4-diensional inner indices, $d=4$,
    and therefore the dimension of each one of the four outer
    indices of a target tensor $T$ is $D=16$ or ${\rm dim}(T)=[16,16,16,16]$.
  }
  \label{fig:noise_dependence}
\end{figure*}

\begin{figure}[!t]
	\centering
  \includegraphics[width=0.6\textwidth]{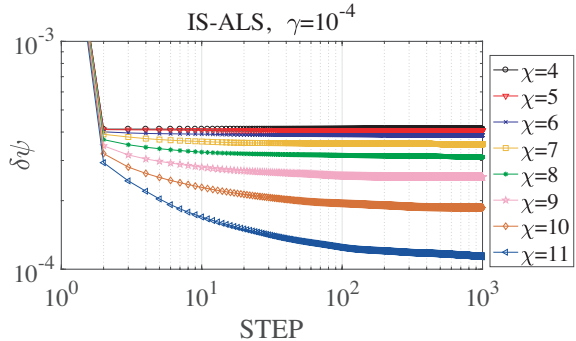}
  \caption{ The dynamics of the approximation error $\delta\psi$ obtained
    by IS-ALS for the same tensor as the one used in Fig.\,\ref{fig:noise_dependence}\,(c).
    Here, the horizontal axis denotes the ALS step, and $\chi$ is the TR-rank or bond dimension. }
  \label{fig:eb_chi}
\end{figure}

Secondly, let us consider the disorder: $\gamma>0$. To see the noise dependence of performances of the IS, IS-ALS and sALS algorithms, we present the dynamics of $\delta\psi$ as a function of iteration step for disorder strengths $\gamma=10^{-6},10^{-5}$ and $10^{-4}$ in Fig.\,\ref{fig:noise_dependence}\,(a), (b) and (c), respectively.
Here, a fourth-order disordered CDL tensor $T$ with $d=4$,
i.e. dim$(T)$ = [16,16,16,16], is used, and the first data\,(at STEP=1)
in the IS-ALS curves are obtained by the IS algorithm.
The accuracy of IS is reduced with the finite random noise compared
to the case of pure CDL tensor.
However, the IS still gives a good TRD and is much more efficient
than the sALS for a given bond dimension\,(see IS-ALS and sALS for $\chi=4$).
In addition, it is improved significantly by a single iteration of ALS,
i.e. about one order of magnitude of $\delta\psi$ is reduced,
as one can see from the second data in the IS-ALS curves.
Note that the order of the approximation error obtained by IS-ALS
is around the order of $\gamma$,
which is good enough for the practical application.
On the contrary, the performance of sALS does not depend much on the disorder strength,
since the special structure of CDL is not taken into account in the algorithm.
Therefore, one expects that the IS does not produce any advantage over the sALS once the
amplitude of the noise exceeds a certain threshold value.
We found that, for tensors with the dimension $[16,16,16,16]$,
the sALS begins to produce a better TRD in the stronger disorder regime
$\gamma \gtrsim 10^{-3}$, above which the CDL structure is almost non-detectable.
Notice that, with the sALS, better accuracy is not guaranteed by a larger
capacity of the ansatz, i.e., a larger $\chi$.
For example, in Fig.\,\ref{fig:noise_dependence}\,(b),
the ansatz with $\chi=5$ is better than the ones with $\chi=6,7$.
Meanwhile, the IS-ALS ensures the better TRD by enlarging $\chi$
as shown in Fig.\,\ref{fig:eb_chi},
in which the same tensor as the one used in Fig.\,\ref{fig:noise_dependence}\,(c)
is decomposed with various $\chi$.
It indicates that the initial latent tensors obtained
by the IS algorithm are already close to the global minimum of the solution.

%

%
\section{Application}
%

\begin{figure}[!t]
	\centering
  \includegraphics[width=0.45\textwidth]{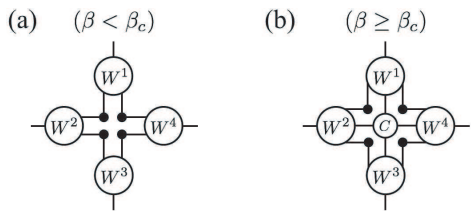}
  \caption{ Schematic figures for the fixed-point tensor of the two dimensional Ising model for (a) $\beta < \beta_c $ and (b) $\beta > \beta_c $ where $\beta\,(\beta_c)$ is the inverse (critical) temperature. }
  \label{fig:ising_fixed_tensors}
\end{figure}

\begin{figure}[!t]
	\centering
  \includegraphics[width=0.5\textwidth]{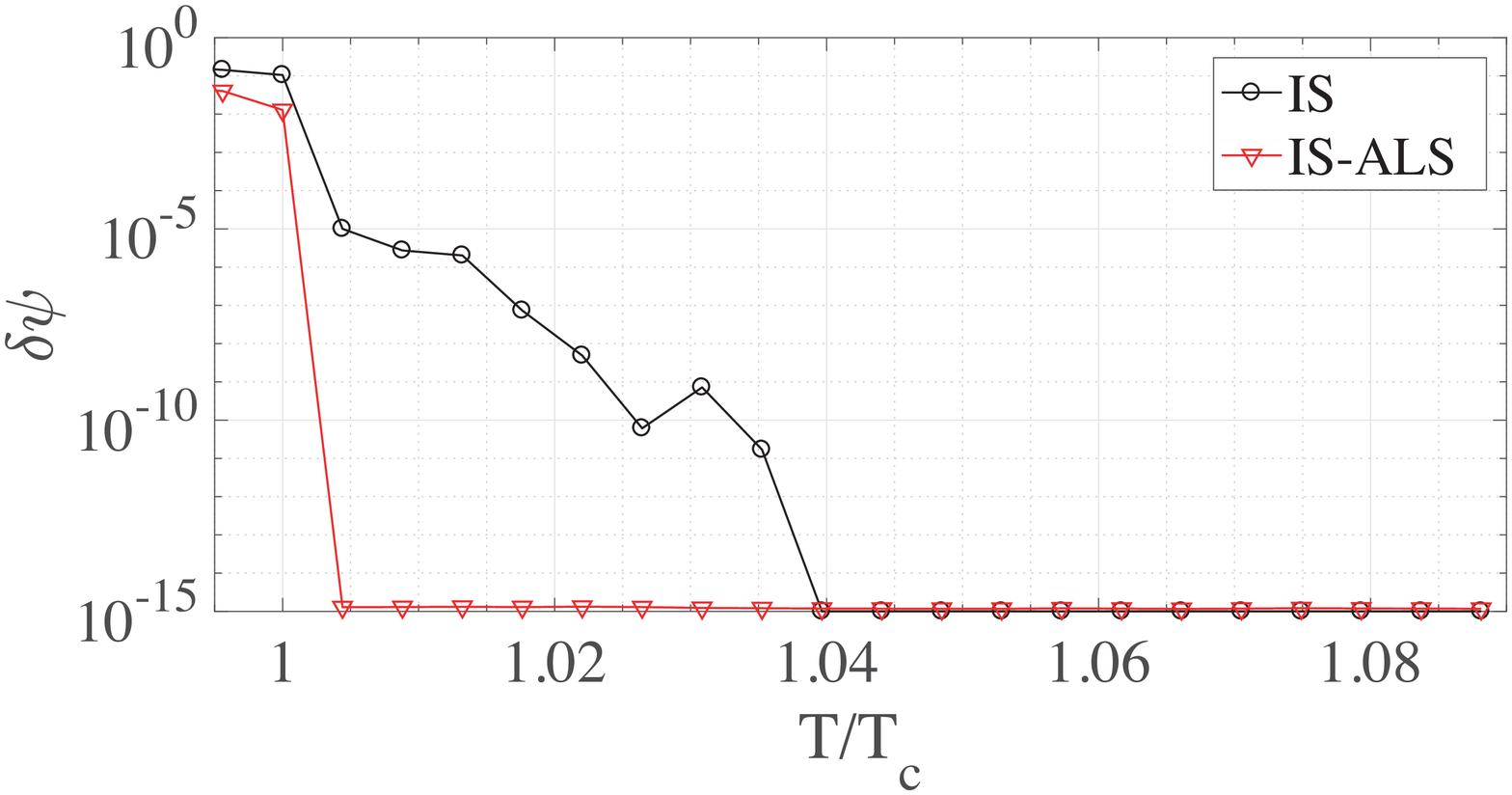}
  \caption{ The approximation error $\delta\psi$ as a function of the temperature\,($T$) by obtained by the IS and IS-ALS with $\chi=4$ for the $[16,16,16,16]$-sized fixed-point tensor of two dimensional Ising model. Here, $T_c$ denotes the critical temperature.}
  \label{fig:ising_tr}
\end{figure}

In this section, we present an exemplary application of IS and IS-ALS algorithms
for tensors obtained from a practical problem; two-dimensional classical
Ising model in the square lattice, whose partition function reads

\begin{eqnarray}
	Z = \sum_{\sigma_1 = \pm1}\cdots \sum_{\sigma_N = \pm1}e^{ \beta\sum_{\langle i,j\rangle} \sigma_i \sigma_j }.
\end{eqnarray}
Here, $N$ is the total number of Ising spin,
$\beta$ is the inverse temperature,
$\sigma_i$ denotes Ising spin at site $i$ and $\langle i,j\rangle$
stands for the nearest-neighbor pair in the whole lattice.
Evaluation of above partition function can be carried out
by contracting a square-lattice tensor network with a tensor
$T^{\rm Ising}_{ijkl}$\,($i,j,k,l=0\,\,{\rm or}\,\,1$) given by\cite{gu09}
\begin{eqnarray}
	&T^{\rm Ising}_{0101} = e^{-4\beta}, \quad
	T^{\rm Ising}_{1010} = e^{-4\beta}, \nn
	&T^{\rm Ising}_{0000} = e^{4\beta},	\quad
	T^{\rm Ising}_{1111} = e^{4\beta}, \quad
	{\rm others} = 0.
\end{eqnarray}
As mentioned in the introduction part,
the TRG of the above tensor flows into the CDL-type tensor at all temperatures.
To be more specific, above the critical temperature or $\beta < \beta_c$,
the fixed-point tensor has the exact CDL structure as depicted
in Fig.\,\ref{fig:ising_fixed_tensors}\,(a).
Therefore, one may expect that the fixed-point tensors can be decomposed very well
by the IS and IS-ALS algorithms.
Figure\,\ref{fig:ising_tr} shows the approximation error
as a function of the temperature\,($T=1/\beta$) obtained
by the IS and IS-ALS with $\chi=4$ for the fixed-point tensor of
dimension $[16,16,16,16]$.
The pure IS decomposition becomes less accurate as the temperature approaches
to the critical point.
However, the following ALS algorithm suppresses the error significantly
and produces the exact TR ansatz\,(red triangle in Fig.\,\ref{fig:ising_tr}).
In other words, one can obtain the exact TR decomposition
at all temperatures above $T_c$ with the IS-ALS.
We believe that this result may open new possibilities
for the {\it index-splitting} based algorithm eliminating
the short-range entanglements in the tensor renormalization group.


At $T \leq T_c$, even the IS-ALS cannot produce a good TRD.
That is because the fixed-point tensors below $T_c$ are not precisely CDL
but have a different structure in which an inner tensor\,($c$-tensor)
is covered by the CDL structure as depicted in Fig.\,\ref{fig:ising_fixed_tensors}\,(b).
However, the index splitting of such {\it generalized}-CDL tensors are
out of the current article. We leave a generalization of the current algorithm,
which applies to the generalized-CDL tensors, for near-future work.

\section{Conclusion}
The CDL structure naturally arises in the TRG procedure.
We have proposed a method, i.e., the index splitting, for efficiently compute
the TRD of a tensor with approximate CDL structure,
by decomposing each index into sub-indices
so that the CDL structure is obvious.
We have demonstrated that the CDL tensor causes serious
convergence problem in previously proposed heuristic algorithms
such as the ALS algorithm,
whereas the present algorithm yields an optimal decomposition
without convergence problem. 

We have also presented that the IS combined with ALS shows great accuracy,
efficiency and stability improvements in TRD for the weakly disordered CDL tensors.
It implies that the index splitting procedure has not only great potential to be applicable to various problems, such as TRG, but also provides us a better understanding of the tensors
we encounter in physics and mathematics.

\appendix
\section{Gauge Fixing}
In this appendix, we discuss the problem of fixing the gauge.
We can (and have to) assume that there are parameters $p,q$ and $r$ and
functions $I(q,r), J(r,p)$ and $K(p,q)$ such that
$t_{ijk}$ is non-vanishing only when $i=I(q,r)$, $j=J(r,p)$ and $k=K(p,q)$
for some combination of $p,q$ and $r$.
We also assume that $I,J$ and $K$ are injective, e.g.,
if $I(q,r) = I(q',r')$ then $q=q'$ and $r=r'$.
Because of the injectivity, we can define the ``inverse'' function
$q_I(i)$ and $r_I(i)$ so that $i = I(q_I(i),r_I(i))$.
However, we do not assume that we know these functions.
Let us define the symbol $\epsilon$ as 
\begin{equation}
  \epsilon_{ijk} \equiv t_{ijk}/|t_{ijk}| \quad (t_{ijk}\ne 0),
\end{equation}
and $\epsilon_{ijk} = 0$ if $t_{ijk}=0$.
Our problem is to find the factors $\alpha_i, \beta_j$ and $\gamma_k$, such that
\begin{equation}
  \epsilon_{ijk} = \alpha_i \beta_j \gamma_k \label{eq:GaugeFactor}
\end{equation}
whenever the left-hand side is not zero.
Then, it is convenient to introduce abbreviation
$\alpha_{qr} \equiv \alpha_{I(qr)}$,
$\beta_{rp} \equiv \beta_{J(rp)}$, and
$\gamma_{pq} \equiv \gamma_{K(pq)}$.
In addition, we define $j_p$ as the ``projection'' of $j$ onto the $p=1$ plane,
i.e., $j_p \equiv J(r_J(j),1)$.
Symbols $j_r, k_p, k_q, i_q, i_r$ are defined similarly.
Assuming the existence of the solution to (\ref{eq:GaugeFactor}),
we have
\begin{equation}
  \epsilon_{I(qr)J(rp)K(pq)} = \alpha_{qr} \beta_{rp} \gamma_{pq},
\end{equation}

With these definitions, let us consider the following functions
\begin{eqnarray}
  \epsilon'_{ijk} & \equiv \bar\epsilon_{ij_pk_p} \epsilon_{ijk}, \label{eq:One} \\
  \epsilon''_{ijk} & \equiv \bar\epsilon'_{i_qjk_q} \epsilon'_{ijk}, \label{eq:Two} \\
  \epsilon'''_{ijk} & \equiv \bar\epsilon''_{i_rj_rk} \epsilon''_{ijk}. \label{eq:Three}
\end{eqnarray}
where $\bar\epsilon \equiv \epsilon^{-1}$.
We can show that $\epsilon'''_{ijk}=1$ for all
$ijk$ that makes $\epsilon_{ijk}$ non-zero.
We can see this as follows
\begin{eqnarray*}
  \epsilon'_{ijk} 
  & = & (\bar\alpha_{qr}\bar\beta_{r1}\bar\gamma_{1q})(\alpha_{qr}\beta_{rp}\gamma_{pq})
  =  \bar\beta_{r1}\bar\gamma_{1q}\beta_{rp}\gamma_{pq} \\
  \epsilon''_{ijk} 
  & = & (\beta_{r1}\gamma_{11}\bar\beta_{rp}\bar\gamma_{p1})
     (\bar\beta_{r1}\bar\gamma_{1q}\beta_{rp}\gamma_{pq}) 
  =  \gamma_{11}\bar\gamma_{p1}\bar\gamma_{1q}\gamma_{pq} \\
  \epsilon'''_{ijk} 
  & = & (\bar\gamma_{11}\gamma_{p1}\gamma_{1q}\bar\gamma_{pq}) 
  (\gamma_{11}\bar\gamma_{p1}\bar\gamma_{1q}\gamma_{pq})
  = 1
\end{eqnarray*}
Therefore, using (\ref{eq:One},\ref{eq:Two},\ref{eq:Three}),
\begin{equation}
  \epsilon_{ijk} = \epsilon''_{i_rj_rk} \epsilon'_{i_qjk_q} \epsilon_{ij_pk_p}
\end{equation}
Since $\epsilon_{ij_pk_p}$, $\epsilon'_{i_qjk_q}$, and $\epsilon''_{i_rj_rk}$
depend only on $i,j$ and $k$, respectively,
we can re-express them as
\begin{eqnarray}
  \alpha'_i \equiv \epsilon_{ij_pk_p}, \quad
  \beta'_j  \equiv \epsilon'_{i_qjk_q}, \quad \mbox{and} \quad
  \gamma'_k \equiv \epsilon''_{i_rj_rk},
\end{eqnarray}
which allows us to write
\begin{equation}
  \epsilon_{ijk} = \alpha'_i \beta'_j \gamma'_k.
\end{equation}
Though this may not be the same as (\ref{eq:GaugeFactor}),
it is a solution to our problem nonetheless.
(Note that the solution is not unique because of the
degree of freedom of multiplying $\alpha'_i$ and
$\beta'_j$ by the same arbitrary factor $\eta_r$,
and similar degrees of freedom concerning $\beta'$ and $\gamma'$,
and $\gamma'$ and $\alpha'$.)

Now, we have to ask if we can compute $\alpha'$, $\beta'$, $\gamma'$
even if we do not know the explicit form of the functions
such as $J(r,p)$ and $r_J(j)$ that we have used,
through $j_p$ and $k_p$, in the definition of $\epsilon'$ and $\epsilon''$.
To answer this question, notice that the element with $p=1$ is largest among
the ones with the same $q$ and $r$.
Sharing the same $q$ and $r$ means sharing the same $i$.
Therefore, 
\begin{equation}
  (j_p,k_p) = \argmax_{(j',k')} |t_{I(j,k)j'k'}|
  \label{eq:Projection}
\end{equation}
where $I(j,k)$ is the value of $i$ that makes $t_{ijk}$ non-zero.
Such $i$ is unique when the tensor is exactly a CDL.
For tensors that are only approximately CDL,
the uniqueness is not guaranteed.
Therefore, we had better define the function $I(j,k)$
by the largest element in order to make the procedure applicable
to such cases:
\begin{eqnarray}
  I(j,k) \equiv \argmax_{i} |t_{ijk}|. 
\end{eqnarray}
With this definition and (\ref{eq:Projection}) together with
(\ref{eq:One}),(\ref{eq:Two}), and (\ref{eq:Three})
we can compute $\alpha'$, $\beta'$ and $\gamma'$
without knowing the explicit form of subindex decomposition.
\ \\

{\bf Acknowledgements}\\
The present work was motivated and inspired by discussions
with S.~Morita, T.~Okubo, T.~Suzuki and K.~Harada.
The computation in the present work is executed on computers at the Supercomputer Center,
ISSP, University of Tokyo, and also on K-computer (project-ID: hp170262).
The present work is financially supported by the MEXT project
``Exploratory Challenge on Post-K computer"\,(Frontiers of Basic Science: Challenging the Limits),
and also by ImPACT Program of Council for Science,
Technology and Innovation (Cabinet Office, Government of Japan).

\Urlmuskip=0mu plus 1mu\relax
\bibliographystyle{elsarticle-num}
\bibliography{references}

\end{document}